\documentclass[fleqn,twoside]{article}
\usepackage{espcrc2}

\usepackage{graphicx}
\usepackage[figuresright]{rotating}

\newcommand{\AmS}{{\protect\the\textfont2
  A\kern-.1667em\lower.5ex\hbox{M}\kern-.125emS}}

\def\({\left(}
\def\){\right)}
\def\[{\left[}
\def\]{\right]}

\title{SUSY dark matter with non-universal gaugino masses}

\author{Andreas Birkedal-Hansen\address{Theoretical Physics Group \\
      Ernest Orlando Lawrence Berkeley National Laboratory \\
      University of California, Berkeley, California 94720 \\
      and \\
      Department of Physics \\
      University of California, Berkeley, California 94720}
        \thanks{This work was supported in part by the DOE 
Contract DE-AC03-76SF00098  and in part by the NSF grant PHY-00988-40.}}

\begin{document}
\begin{titlepage}
\begin{center}
      \hfill  LBNL-50109 \\
      \hfill  UCB-PTH-02/15 \\
      \hfill hep-ph/0204176 \\
\hfill April 2002
\vskip .5in

{\large \bf SUSY dark matter with non-universal gaugino masses }\footnote{Talk presented at Dark Matter 2002, February 20-22, 2002,
Marine del Rey, CA, to be published in the 
proceedings.}\footnote{This work was supported in part by the DOE 
Contract DE-AC03-76SF00098  and in part by the NSF grant PHY-00988-40.}\\[.1in]

Andreas Birkedal-Hansen

{\em Department of Physics,University of California, and

 Theoretical Physics Group, 50A-5101, Lawrence Berkeley National Laboratory,
   Berkeley, CA 94720, USA}

\vspace{2pc}
{\bf Abstract}
\end{center}

In this talk we investigate the dark matter prospects for supersymmetric models with non-universal gaugino masses.  We motivate the use of non-universal gaugino masses from several directions, including problems with the current favorite scenario, the cMSSM.  We then display new corridors of parameter space that allow an acceptable dark matter relic density once gaugino mass universality is relaxed.  We finish with a specific string-derived model that allows this universality relaxation and then use the dark matter constraint to make specific statements about the hidden sector of the model.
\vspace{1pc}

\end{titlepage}

\begin{abstract}
 In this talk we investigate the dark matter prospects for supersymmetric models with non-universal gaugino masses.  We motivate the use of non-universal gaugino masses from several directions, including problems with the current favorite scenario, the cMSSM.  We then display new corridors of parameter space that allow an acceptable dark matter relic density once gaugino mass universality is relaxed.  We finish with a specific string-derived model that allows this universality relaxation and then use the dark matter constraint to make specific statements about the hidden sector of the model.
\vspace{1pc}
\end{abstract}

\maketitle

\section{Introduction}



This talk summarizes recent work \cite{Andreas&Brent} investigating the effects of gaugino mass non-universality on dark matter prospects for low-energy models of broken supersymmetry.  When we allow this non-universality in the bino and wino masses, significant new regions of parameter space now satisfy the dark matter constraint.  We show that these regions are largely independent of the common scalar mass.

This freedom in gaugino masses is highly motivated, especially from string-derived models \cite{BGWI,Ibanezetc}.  When we look at a specific class of models allowing freedom in the bino and wino masses (specifically the ratio $M_{2}/M_{1}$), we see that the dark matter constraint yields interesting results.  We find that the coupling of the dark matter constraint with a few basic phenomenological constraints allows us to determine the matter and gauge content of the hidden sector of the theory, assuming only that the matter comes in fundamental representations.

The cMSSM is by far the most studied supersymmetric model in the literature.  In the cMSSM, one finds that the neutralino relic density in most of the $\left(M_{0},M_{1/2}\right)$ parameter space is determined in large part by annihilation to a light fermion anti-fermion pair.  Since the preferred dark matter candidate, the lightest neutralino, is mostly bino in character, the interaction strength is rather weak.  These two facts coupled together lead to a neutralino relic density that is too large over most of the $\left(M_{0},M_{1/2}\right)$ plane, as shown in Figure 1.  

Here the region with the cosmologically preferred relic density is the one bounded by two solid lines, containing the points labelled $A$ and $B$.  The regions containing points $C$ and $D$ have a relic density that is too high, conflicting with measurements of the age of the universe.  The near-vertical dotted lines denote contours of constant Higgs mass.  It is apparent that the current limit of $m_{h}\geq 114.1$ GeV \cite{Higgslimit} removes large sections of parameter space, leaving only the Higgs pole (the vertical region) and the neutralino-stau coannihilation channel (near-horizontal region).  Both of these regions are aesthetically unappealing since they require a very tight mass relation between a fermion and an unrelated scalar.    

In the cMSSM, the mass of the lightest neutralino is determined by the bino mass parameter, $M_{1}$, over most of parameter space.  This is due to the constraint of $M_{2}\simeq 2 M_{1}$ (at the weak scale) coming from the GUT relation on the gauge couplings.  However, once $M_{2}$ is allowed to take a value unrelated to $M_{1}$, the mass of the lightest neutralino is generally determined by a mixture of $M_{2}$ and $M_{1}$.  This can be seen from the neutralino mass matrix:

\begin{figure}
\includegraphics[scale=.4]{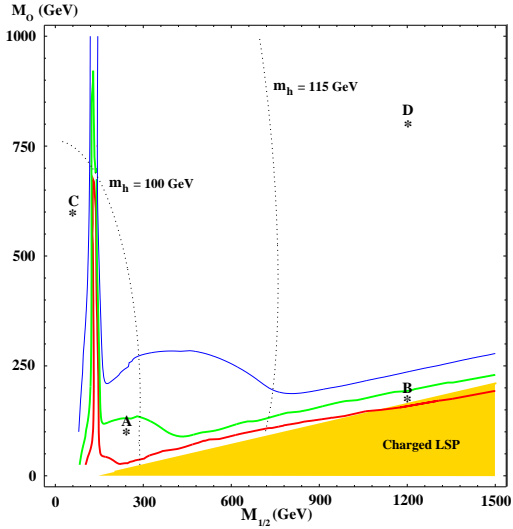}
 \caption{{\small { Preferred Dark Matter Region
                for cMSSM with $\tan \beta = 3$}. Contours of $\Omega_{\chi}{\rm h}^2$ of 0.1 (bottom-most contour), 0.3 and
              1.0 (top-most contour) are given. The shaded region is ruled out
              by virtue of having the stau as the LSP. We have also added Higgs mass contours of $m_{h} = 100$ GeV and $m_{h}=115$
              GeV. 
             }}
\end{figure}

{\footnotesize\begin{equation}
\(\begin{array}{cccc}M_{1} & 0 & -s\, c_{\beta} M_{Z} &
    s\, s_{\beta} M_{Z} \\ 0 & M_{2} & c\, c_{\beta} M_{Z}& - c\, s_{\beta} M_{Z}\\ - s\,
    c_{\beta} M_{Z} & c\, c_{\beta} M_{Z} & 0 & -\mu \\ s\, s_{\beta} M_{Z} & -c\, s_{\beta} M_{Z} &
    -\mu & 0 \end{array}\)
\end{equation}}

Where $s=\sin \theta_{W}$, $c=\cos \theta_{W}$, $s_{\beta}=\sin \beta$ and $c_{\beta}=\cos \beta$ with $\tan \beta$ the ratio of the Higgs {\it vev}s.  In Figure 2, we show the dependence of the relic density on $M_{2}/M_{1}$ evaluated at the GUT scale.  The near-vertical lines are contours of constant relic density, having values (starting from the left) of: $\Omega h^2 = 0.01, 0.1, 0.3, 1.0, 10.0$.  In this plot we have picked the value of $M_{1/2}=min\left(M_{2},M_{1}\right)$ to be $200$ GeV, but the story remains much the same for values of $M_{1/2}$ in the range of hundreds of GeV.  We can see that the requirement of proper relic density prefers a value of $M_{2}/M_{1}$ quite different from the cMSSM value of 1.  

\begin{figure}
\includegraphics[scale=.4]{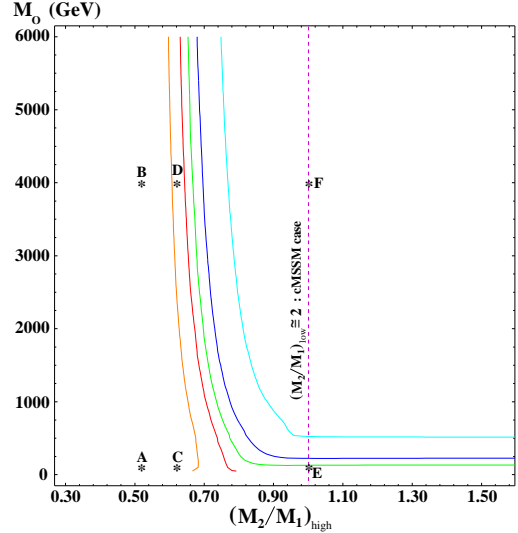}
\caption{{\small { Preferred Dark Matter Region
                for Non-universal Gaugino Masses}. Contours of $\Omega_{\chi}
              {\rm h}^2$ of 0.01, 0.1, 0.3, 1.0 and 10.0 from left to
              right, respectively, are given as a function of $M_{2}/M_{1}$ at
              the high scale.
             }}
\end{figure}

We can also see that the requirement of proper relic density is largely independent of the scalar mass $M_{0}$.  This, again, is in stark contrast to the situation in the cMSSM.  In the cosmologically preferred regions with non-universal gaugino masses,  the relic density is not largely determined by annihilation into fermion anti-fermion pairs through squark exchange.  Another feature of lowering $M_{2}/M_{1}$ is the increasing degeneracy of the masses of the lightest and next-to-lightest neutralinos.  Additionally, the mass of the lightest chargino also approaches the mass of the lightest neutralino for values of $M_{2}/M_{1}$ below the cMSSM value.  The reason this happens at a value of $M_{2}/M_{1}<1$ in Figure 2 is just due to RGE running of the gaugino masses from the GUT scale (as in Figure 2) down to the weak scale.  The masses of the next-to-lightest neutralino and the chargino are usually determined by $M_{2}$, so one would naturally expect their masses to approach the lightest neutralino mass if the aforementioned mass ratio is lowered.  Now coannihilation channels between two different neutralinos and also between neutralinos and charginos become important.

The BGW models\cite{BGWI} are explicit examples of string-derived models that exhibit this freedom in $M_{2}/M_{1}$.  The BGW models are models of modular invariant supergravity derived from the weakly coupled heterotic string.  Supersymmetry is broken non-perturbatively, through gaugino condensation in a hidden sector.  The superpartner spectrum is determined by the dynamics of the condensing gauge group with the largest beta function.  The phenomenology of the low-energy spectrum has been studied in \cite{BGWphenI}.  The hidden sector matter content and beta function determine the parameters $b^{\alpha}_{+ eff}$ and $b_{+}$.  The string-scale unified gauge coupling is determined by the vacuum expectation value of the dilaton.  The gaugino masses are given by:

\begin{equation}
M_{a} = \frac{g_{a}^{2}}{2}\(\frac{3b_{+} \(1+b'_{a} l\)}{1+b_{+} l} -3 b_{a}\) M_{3/2}
\end{equation}
where $g_{a}$ is the gauge coupling evaluated at the condensation scale, $b_{a}$ is the corresponding beta function coefficient and $b'_{a}$ is an anomaly matching coefficient.  As we can see, varying the hidden sector parameter $b_{+}$ can indeed vary the ratio $M_{2}/M_{1}$.  In these models the scalar masses begin at a unified value, just as in the cMSSM.  Thus, we can plot the cosmologically preferred region in the $\left(M_{0},b_{+}\right)$ plane, as shown in Figure 3.  The lines of constant $\Omega h^2$ are given by the solid lines that are nearly vertical at the top of the plot, with values $0.01, 0.1, 0.3, 1.0, 10.0$ (starting from the left).  The shaded region is excluded by limits on the gluino and chargino masses ($190$ and $90$ GeV respectively).  We can see, then, that the cosmologically preferred region is almost indepedent of scalar mass, as long as the scalar mass is sufficiently heavy to pass the gluino and chargino limits.  This agrees with the earlier general analysis of models with $M_{2}/M_{1}$ freedom.

\begin{figure}
\includegraphics[scale=.45]{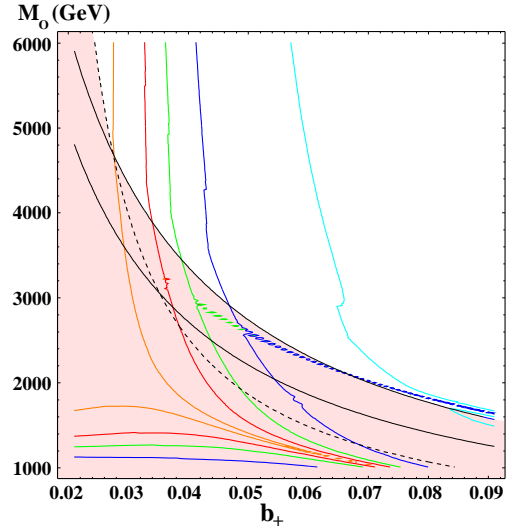}
 \caption{{\small { Preferred Region in the BGW
                Model}. Contours of $\Omega_{\chi} {\rm h}^2$ are given as a function of $M_{0}$ and $b_{+}$
              by the solid lines. The shaded region is
              excluded by the gluino mass and chargino mass limits.}}
\end{figure}

\begin{figure}
\includegraphics[scale=.45]{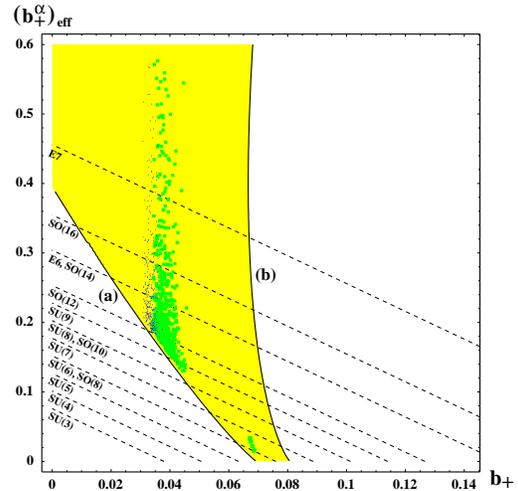}
\caption{{\small { Preferred Dark Matter Region in
                Hidden Sector Configuration Space.}The fine
              points on the left have the preferred value $0.1 \leq
              \Omega_{\chi} {\rm h}^2 \leq 0.3$ and the coarse
              points have $0.3 < \Omega_{\chi}{\rm h}^2 \leq 1.0$.}}
\end{figure}

If we only look at the hidden sector parameters, $b^{\alpha}_{+ eff}$ and $b_{+}$, we can see which hidden sector gauge groups are allowed.  In Figure 4, the fine dots represent cosmologically allowed neutralino relic densities.  The coarse dots represent values the are slightly too large, with $\Omega h^2$ being between $0.3$ and $1.0$.  The swath bounded by lines (a) and (b) is the region in which the hidden sector Yukawa couplings and gravitino mass have reasonable values of $10\geq \(c_{\alpha}\)_{eff}\geq 0.1$ and $100$ GeV $\leq M_{3/2} \leq 10$ TeV.  In the $\left(b_{+}, b^{\alpha}_{+ eff}\right)$ plane, a given gauge group makes a slanted line.  

We can see that the only gauge groups allowed by the dark matter constraint are $SO\left(12\right)$, $SO\left(14\right)$, $SO\left(16\right)$, $E_{6}$, and $E_{7}$.  This is a non-trivial constraint on the largest condensing gauge group of the hidden sector.  However, we should note that we can also allow other forms of dark matter to contribute to the total density.  In Figure 4, this would allow any point in the shaded region to the left of the cosmologically preferred points to also be viable.  If we again require $0.1\leq \Omega h^2\leq 0.3$ and also assume that hidden sector matter only exists in fundamental representations (vector fundamentals for $SO\(2n\)$), then we can also rule out $E_{7}$.  This is because a gauge invariant matter condensate cannot be formed that allows the neutralino relic density to have roughly the preferred value.  In fact, $E_{6}$ also predicts a relic density that is a little too high ($\Omega h^2 = 0.63$), and $SO\left(12\right)$ and $SO\left(14\right)$ predict relic densities that are a little below the preferred region ($\Omega h^2 = 0.076, 0.069$ respectively).  However, with $SO\left(16\right)$ we find a relic density of $\Omega h^2 = 0.194$.  If the matter in the hidden sector exists only in these fundamental representations, then the only cosmologically preferred gauge group is $SO\left(16\right)$.  Even though hidden sector matter is not required to only exist in fundamental representations, we still find that dark matter constrains the largest hidden sector gauge group to the list: $SO\left(12\right)$, $SO\left(14\right)$, $SO\left(16\right)$, $E_{6}$, and $E_{7}$.

In summary, the prospects for viable dark matter in the cMSSM at low $tan \beta$ are becoming more limited, due both from theory and experiment.  Many highly motivated supersymmetric models contain freedom to allow non-universal gaugino masses, a freedom not present in the cMSSM.  Exploiting this freedom allows significantly more room for viable dark matter.  Dark matter with non-universal gaugino masses does not require a curious relationship between the lightest neutralino and an unrelated scalar to find an allowed relic density.  Freedom in $M_{2}/M_{1}$ also allows much heavier scalar masses to be generically compatible with dark matter than in the cMSSM.  The dark matter constraint can be used to significantly limit specific models that allow non-universal gaugino masses.  As an example, we chose models of modular invariant supergravity and showed that the dark matter constraint can limit the allowed configurations of the hidden sector to a few gauge groups.

\section*{Acknowledgements}

This work was performed in collaboration with Brent D. Nelson.

\end{document}